\documentclass[pdflatex,sn-mathphys-num]{sn-jnl}

\usepackage{subcaption} 
\usepackage{graphicx}%
\usepackage{multirow}%
\usepackage{amsmath,amssymb,amsfonts}%
\usepackage{amsthm}%
\usepackage{mathrsfs}%
\usepackage[title]{appendix}%
\usepackage{xcolor}%
\usepackage{textcomp}%
\usepackage{manyfoot}%
\usepackage{booktabs}%
\usepackage{algorithm}%
\usepackage{algorithmicx}%
\usepackage{algpseudocode}%
\usepackage{listings}%
\usepackage{overpic}
\usepackage{nameref}

\usepackage{changes}

\theoremstyle{thmstyleone}%
\theoremstyle{thmstyletwo}%

\theoremstyle{thmstylethree}%

\begin{document}

\title[]{Verification of the Outer Space Treaty with Cosmic Protons}


\author*[]{\fnm{Areg} \sur{Danagoulian}}\email{aregjan@mit.edu}


\affil[]{\orgdiv{Department of Nuclear Science and Engineering}, \orgname{Massachusetts Institute of Technology}, \orgaddress{\street{77 Massachusetts Avenue}, \city{Cambridge}, \postcode{02139}, \state{Massachusetts}, \country{United States}}}



\abstract{The Outer Space Treaty (OST) was opened to signatures in 1967, and since then 117 countries, including China, the United States, Russia, have become part of it~\cite{OST}. Among other stipulations the treaty bans the placement of nuclear weapons in outer space.  Recently the US government has raised worries that Russia is testing nuclear-armed anti-satellite weapon components, with the possibility that it will place a nuclear weapon in space. Such a device, if detonated, would destroy most of the satellites in the Low Earth Orbit.  This danger is compounded by the lack of a verification mechanism for the OST. No methodologies of verification have been proposed in the open peer reviewed literature. This study presents a concept and a feasibility study for verifying a satellite's compliance to the OST by observing the neutrons induced by spallation from the $\sim$GeV protons in the inner Van Allen radiation belts~\cite{carpenter1977pulsed}.  The calculations show that a 9U CubeSat sized detection platform can identify a thermonuclear weapon from the distance of 4~km in approximately one week of observation.  This conceptual study will stimulate and inform future research and development of verification platforms for OST.}




\maketitle

On February 2, 2022 Russia launched the Kosmos2553 satellite into Low Earth Orbit (LEO) at an altitude of approximately 2000 km. Russia claims that the satellite is part of its Neitron Radar System and is to be used for surveillance and remote sensing~\cite{rfe}. The US officials have expressed worries that Kosmos2553 is a test platform for an anti-satellite weapon (ASAT). The US asserted that the satellite may contain nuclear weapon components undergoing in-space qualification as part of a systematic effort by the Russian Federation to develop a nuclear-charge-based ASAT, "which, if detonated, could potentially wipe out an entire orbit of assets"~\cite{vipin}.  In fact, the Starfish Prime test of 1962, which detonated a 1.4 megaton thermonuclear warhead in space, injected 10$^{29}$ electrons into the inner Van Allen belt, increasing its electron population by several orders of magnitude and destroying many of the early artificial satellites of the era~\cite{stassinopoulos2015starfish}.  Such a development would constitute a major violation of the OST, which was signed and ratified by the US, China, and Russia, along with 114 signatories bans the placement of nuclear weapons in orbit or in outer space.   The urgency of this matter became particularly clear in a recent White House executive order, which clarified the importance of  "ensuring the ability to detect, characterize, and counter threats to United States space interests from very low-Earth orbit and through cislunar space, including any placement of nuclear weapons in space"~\cite{ExecOrder}. While the Outer Space Treaty is almost sixty years old, it has always lacked robust means of verification for space-based nuclear threats~\cite{porteous}. Since then, nuclear detection technologies have advanced significantly.   This study proposes a novel concept of verification.  This is a major technical challenge; LEO poses a very harsh radiation environment in which traditional nuclear detection methods are encumbered by the bombardment of protons and electrons trapped in the inner Van Allen Radiation Belt. This study considers a hypothetical satellite in the same orbit as Kosmos2553, which contains a thermonuclear weapon.  It proposes a conceptual design of a directional neutron detector package the size of a 9U CubeSat, carried by an inspecting satellite, henceforth referred to as the inspector. The results show that it can detect the neutrons produced by proton induced spallation on the thermonuclear device's uranium radiation case as a way of confirming that the suspected satellite does carry a thermonuclear device. 
The advantage of the approach described in this study is entirely relies on the naturally present radiation of the inner Van Allen radiation belt for interrogating the suspect satellite.

\section*{Van Allen belt particles and neutron source}

The inner Van Allen Radiation belts consist of trapped electrons of $\sim$MeV and trapped protons of MeV-GeV energies.  These electrons and protons are thought to be created as a result of the Cosmic Ray Albedo Neutron Decay (CRAND) process.  GeV-TeV Galactic Cosmic Rays (GCR) hitting Earth’s upper atmosphere produce hadronic showers and hadronic cascades on oxygen and nitrogen nuclei.  These showers result in the production of mesons, muons, neutrons, and other particles.  While many of these cosmogenic particles are emitted downward and reach the Earth's atmosphere, some of the neutrons among them are emitted radially upward.  While most of these neutrons escape Earth's orbit, a very small fraction of them undergo beta decay within the Van Allen belts due to their 880~s lifetime. The resulting electrons and protons can then be trapped within the belts and reside there for a long time.  Because of this, the proton populations can be seen as a long-term integral of cosmic ray activity and can be thought of as a "fossil record" of galactic and extragalactic cosmic activity, with little dependence on the sun's activity or other fast-changing factors. The $>200~$MeV proton population, in particular, is of interest:  it peaks at around the L-shells with McIlwain L values of L=1.3-1.6~\cite{mazur2023relativistic,johnston2015recent}.  These shells intersect Earth's magnetic equatorial plane at altitudes of approximately 1800-3700 km. 

\begin{figure}[p]
    \centering
    \includegraphics[width=\linewidth]{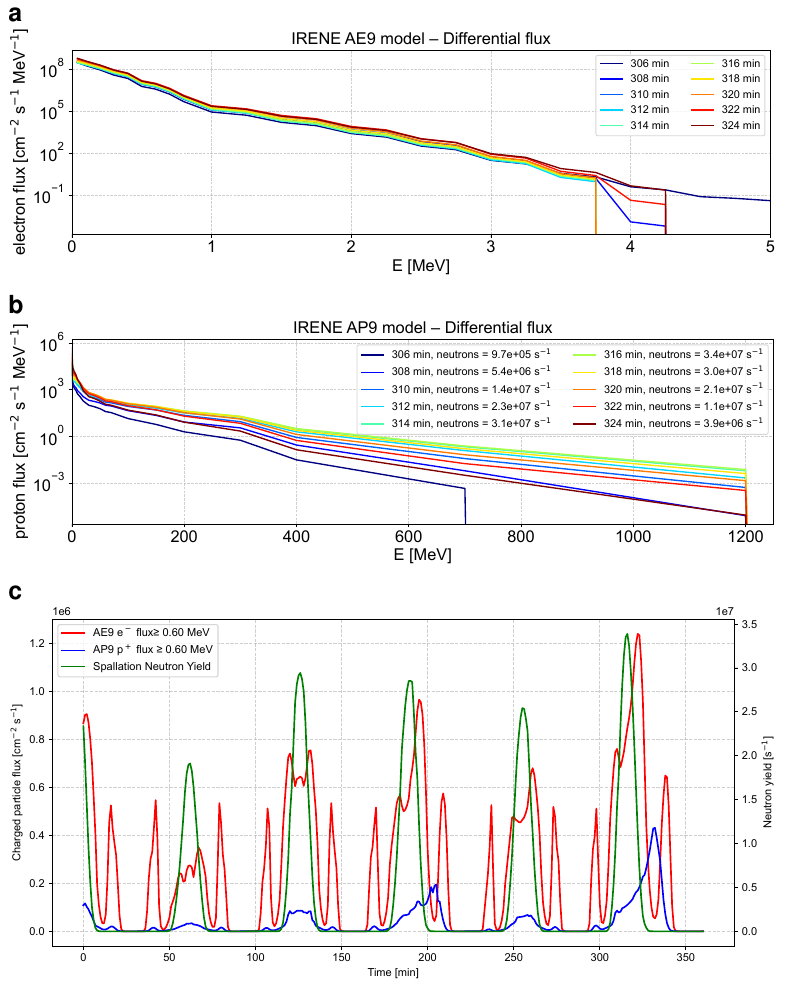}
    \caption{Particle spectra, fluxes, and spallation neutron yield along the Kosmos2553 orbit. 
    (a) Electron spectra around minute 316.
    (b) Proton spectra around minute 316.
    (c) Proton flux, electron flux, and spallation neutron yields vs. time over three orbits.}
    \label{fig:time}
\end{figure}

In this study, we consider the orbit of Kosmos2553, which, at the time of this writing, has a perigee of 2000.3 km, an apogee of 2006.5 km, a period of 127.1 minutes, and an inclination of 67.1$^\circ$. The IRENE AP9 and AE9 v1.58 models were used to determine the spectra of electrons and protons at a one-minute resolution along the orbit of the satellite~\cite{IRENE}.  To understand the expected backgrounds and the signal terms, we first observe the flux of the electrons and protons. The dependence of the total flux of these particles for $E\geq0.6$~MeV on time can be seen in Fig.~\ref{fig:time}(c), while the individual electron and proton spectra for a selection of points along the Kosmos2553 orbit are plotted in Figures~\ref{fig:time}(a) and \ref{fig:time}(b), respectively.  

The protons of intermediate $\sim$GeV energy interact with the high-Z material in the hypothetical thermonuclear device in a satellite, producing many neutrons through the process of spallation. The effective proton energies for neutron spallation are $E\gtrsim200$~MeV.  To determine the neutron production rates through spallation, we use the spallation model for thick targets by Carpenter~\cite{carpenter1977pulsed}.  The empirical equation of the neutron yield's dependence on the proton energy for uranium, proposed by Carpenter, is $Y=50(E/$GeV$-0.12)$ for $E\geq0.2$~GeV. This model has shown good agreement with data for thick uranium targets of approximately 95~kg.  In this work, we are guided by the existing estimates indicating that the weight of the thermonuclear charge, primarily driven by its uranium radiation case, is more than 100 kg~\cite{Harvey01081994}. This circumstance makes this simple model a good order-of-magnitude starting point for our calculations of the neutron yield. See Supplementary Note 1 for more detail. 

In order to determine the neutron yield, every bin in the proton energy spectrum above $\geq 200$~MeV is multiplied by the formula above, with the additional assumption of the thermonuclear charge's size of approximately 50$\times$20~cm$^2$.  The resulting neutron production rates, or yields, can be seen as the green line in Fig.~\ref{fig:time}(c).  
Additionally, the plot shows that the majority of electrons arrive at times when no spallation neutrons are produced, allowing for the timing of the data acquisition (DAQ) onboard the inspector platform to filter out most electrons.  There is, however, a significant residual count of electrons at the peak of neutron production:  this will require careful considerations for DAQ operation to suppress the electron interference, as will be discussed later.

The proton flux, while producing the spallation neutron source term that is the basis of detection, can also be a significant contributor to background in the inspector's detector due to the fact that the neutron scintillator of choice, EJ-276, fundamentally detects neutrons by proton recoil and, as such, is intrinsically a proton detector.  A detailed discussion of how these contributions are suppressed via active shielding anti-coincidence methods can be found in the following sections.

\section*{Detector pixel construction}

In the conceptual model presented in this study the detector array consists of two planes  of 30$\times$30 pixels separated by 10~cm.  Each pixel consists of an inner cube of EJ-276 of the size of 0.9$\times$1$\times$1 cm$^3$, covered above and below by 1~mm-thick plates of carbon vapor deposition (CVD) single crystal diamond detectors. The total size of the pixel is then 1.2$\times$1$\times$1~cm$^3$ big. The 1~mm gap between the inner detector and the outer diamond veto was left to account for the presence of silicon photomultiplier (SiPM) arrays, which will read out the light from the inner detector. The dual plane array then approximately amounts to a volume of a 9U CubeSat. A 3D rendering of the simulated geometries of a single pixel can be seen in Fig.~\ref{fig:pixel_planes}(a), with a rendering of the full dual-array detector assembly displayed in Fig.~\ref{fig:pixel_planes}(b). The figure shows the sampling of the protons; for the purposes of the sensitivity and specificity analysis, the MC simulation samples the proton energy from the spectrum for minute 316, as described in Fig.~\ref{fig:time}.

The simulation model also includes a very thin dead layer of 60~nm. 
A detailed discussion on the characteristics of single crystal diamond detectors can be found in a study by Reichelt et al.~\cite{reichelt2025ultra}, which shows that the diamond detectors can have rise times of less than 50~ps and overall response times of less than 0.5~ns -- which is an important consideration when estimating the detection dead times due to exposure to electrons and protons in the radiation belt. Additionally, Ref.~\cite{ogasawara2015single} shows that these detectors can detect protons down to energies of just 14~keV, ensuring that protons with enough energy to traverse the pixel cannot pass through the top and bottom veto layers undetected. 

The outer diamond layer acts as an active veto shielding:  the inner neutron detector and the outer diamond detectors are operated in anti-coincidence.  This is necessary for rejecting all events in which a cosmic proton traverses the pixel.  At the same time, the 1~mm-thick diamond will have a very small effect on the neutrons due to the small probability of interaction of the spallation neutrons with the veto. 

The nuclear reaction considerations motivating this design choice and the Geant4 simulation parameterization are described in Methods.

\section*{Anti-coincidence based p/e\textsuperscript{-} rejection}

As detailed in Methods, the MC simulation of proton anti-coincidence demonstrates that all proton tracks are rejected via the veto, ensuring no proton hits in EJ-276 are mistaken for neutrons (Fig.~\ref{fig:protons_and_theta}(a)).

The detector array, in addition to being exposed to protons, is also exposed to the electrons trapped in the inner Van Allen Radiation belt. The spectra for a selection of minutes along the orbit can be seen in Fig.~\ref{fig:time}(a). 
The MeV electrons , like the protons, are essentially guaranteed to trigger a response in the veto detector with an energy of $E_{\mathrm{veto}}>0.6$~MeV, guaranteeing their rejection through anti-coincidence. The remaining compounding circumstance involves the possibility that the electrons may trigger bremsstrahlung in the diamond detectors and that the resulting $\sim$MeV photons may then trigger EJ-276 inner scintillator in various other pixels in the two arrays.  Additionally, due to their longer ranges in diamond, some electrons with $E_{\mathrm{veto}}<0.6$~MeV may leak into EJ-276 without triggering a veto signal.  In a physical system, both electrons and bremsstrahlung photons can be rejected via the pulse shape discrimination (PSD) technique in the affected pixel's scintillator itself. Finally, the detection system is also exposed to GCRs, which constitute a relatively low flux of highly energetic charged particles. These will be tagged and vetoed out, not unlike the protons described earlier.  

During the neutron production periods, as can be seen in Fig.~\ref{fig:time}(c), the electron flux and the proton flux peak at approximately $8\times10^5$~cm$^{-1}$s$^{-1}$.  This will create dead times, which are accounted for later in the calculation of the neutron signal term. Future research should focus on reducing these high rates, e.g. by placing thin shielding of low-Z hydrogenous materials such as high density polyethylene.

\section*{Directionality-based neutron rejection}\label{sec:backprojection}

The detection of spallation neutrons from a satellite carrying a thermonuclear device is challenging because of a number of backgrounds.  We have discussed the electron and proton fluxes in the Van Allen radiation belts in the LEO, as well as the methods for rejecting them via anti-coincidence veto methods.  However, in addition to these, there are a number of neutron backgrounds which cannot be rejected by simple anti-coincidence methods.  These are: atmospheric albedo neutrons; neutrons from proton spallation or from various (p,n) and (p,p$^\prime$n)reactions in the inspecting satellite and in the diamond veto detectors.

We determine that the total flux of these background neutrons can be significant, necessitating the reconstruction of directional information about the incident neutron as a means of discriminating between the neutrons from the suspected satellite and the background neutrons from the sources listed above.  Such techniques, referred to as neutron scatter cameras, have been studied over the last few decades and have been shown to be quite effective in reconstructing the direction of the incident neutron with an angular resolution of better than 10$^\circ$~\cite{WEINFURTHER2018115,poitrasson2014dual,keefe2022design}.  As described earlier, our study envisions a detection system that consists of two planes of pixels separated by 10~cm.  This system will detect neutrons and record only those events where there is a neutron coincidence between the two planes. Then, by registering the energy deposition of the neutrons in the first plane, the scattering angle as determined by the hit positions in the two planes, as well as the time $\Delta t$, one can do kinematic analysis of the incident neutron and determine whether it arrived from the direction of the suspected satellite.  In other words, the directional information will then be used to differentiate between neutrons coming from above from the suspect satellite located above from all other sources of neutrons which come from other directions. 

Here it is important to note that our approach assumes that the inspector satellite will be positioned vertically beneath the suspected satellite undergoing inspection.  To accomplish this, the inspector will need to be positioned in an orbit with a perigee that is below the suspect's apogee and an apogee that is above the suspect's perigee.  The values of apogee and perigee can be chosen such that the periods of the two satellites match, allowing for an extended inspection period. Additionally, the phase of the inspector will be chosen such that the inspector is below the suspect exactly at a time when the $\sim$GeV proton flux is at a maximum, i.e. when the two traverse the $L\sim1.4$ L-shell. In this way, the inspector will be below the suspect at all times when the spallation neutron signal is at its peak, allowing for a clear 180$^\circ$ separation between neutrons from the suspect above and the atmospheric albedo neutrons from below. The two satellites -- the inspector and the suspect -- will be in a co-orbital configuration where the two objects oscillate around each other and maintain close proximity for a long time with little need for propulsion to maintain formation.  Alternatively, the inspector  can be in the same orbit as the suspect, in a leader-follower configuration with a delay or stand-off of a few kilometers; the directional analysis of the incident neutrons would allow for angular discrimination between horizontally arriving spallation neutrons and vertically arriving albedo neutrons in this scenario as well.  For the purpose of simplicity in the associated analysis, however, this study focuses on the scenario in which the inspector is below the suspect at the time of the main neutron "pulse." A detailed treatment of the relative motion using the linearized Clohessy--Wiltshire (CW) equations can be found in Supplementary Note 0.

The kinematic reconstruction procedure for $\theta_{\mathrm{error}}$ and the MC simulation used to characterize the directional performance are detailed in Methods.
\begin{figure}[p]
    \centering
    \includegraphics[width=1\linewidth, angle=0]{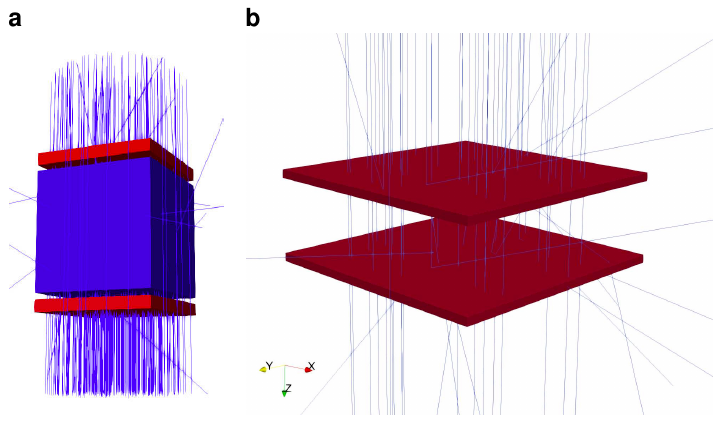}
    \caption{Model of the 9U CubeSat detector. 
    (a) Rendering of the simulation of a single pixel of the array.  The red volumes correspond to the diamond veto detectors. For purposes of clarity, the 60~nm dead layer is omitted and only axial protons are included.
    (b) Rendering of the full system of the two detector planes. The planes are 30$\times$30~cm$^2$, i.e. 30$\times$30 pixels big.}\label{fig:pixel_planes}
\end{figure}

\begin{figure}[p]
    \centering
    \includegraphics[width=\linewidth]{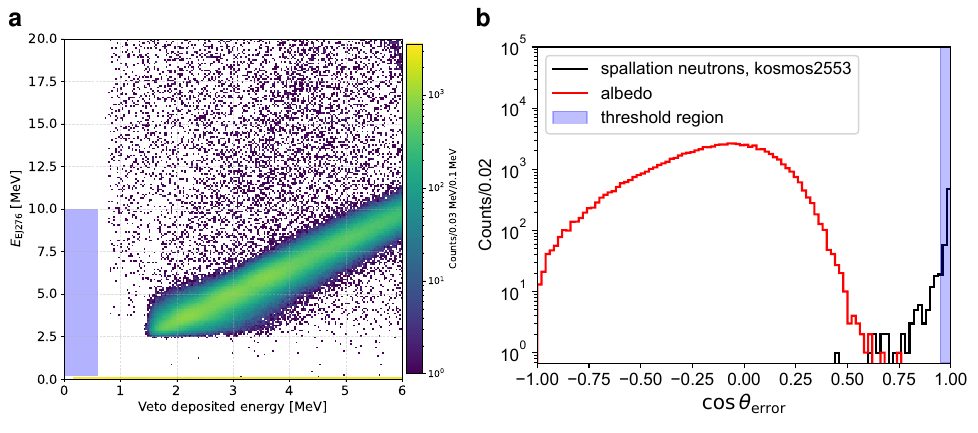}
    \caption{Suppression of proton and neutron backgrounds. (a) A histogram of the distribution of EJ-276 signal vs. veto signal for incident protons. The blue rectangle describes the acceptance cut.  All the proton events are rejected.
    (b) Distribution of $\cos(\theta_{\mathrm{error}})$ for the simulated albedo neutron and the spallation neutron flux. The $\cos(\theta_{\mathrm{error}})>0.95$ threshold cut (blue region) describes the condition for accepting neutrons. It retains 87\% of spallation counts, while rejecting all albedo neutrons. 
    }\label{fig:protons_and_theta}
\end{figure}

A 3D rendering of the simulation and the distribution of $\cos\theta_{ \mathrm{error}}$ can be seen in Fig.~\ref{fig:pixel_planes}(b) and Fig.~\ref{fig:protons_and_theta}(b).  The neutron energies are sampled from the experimental neutron spallation distributions described in Ref.~\cite{carpenter1977pulsed}.  The conservative cut $\cos(\theta_{\mathrm{error}})>0.95$ is applied to eliminate any contributions of atmospheric neutrons. The cut results in the retention of $\approx$87\% of the vertical spallation neutrons that trigger coincidences. The combined intrinsic system coincidence efficiency is determined to be $\epsilon = (0.54\pm0.02)\%$, meaning that 0.54\% of vertical spallation neutrons incident on the detector plane are registered with $\cos(\theta_{\mathrm{error}})>0.95$. The uncertainty in $\epsilon$ is due to MC statistics. An extended simulation was also run to determine the false positive rates with this $\theta_\text{error}$ cut.  It was determined that for a 9U satellite and an observational time of 7.2 days the expected false neutron count is expected to be $<0.011$ counts.   See Supplementary Note 3 for further detail. 

Additionally, the system needs to also reject the spallation neutrons generated by protons in the diamond detectors above the EJ-276 pixels. To simulate this effect, we generate neutrons isotropically in one of the diamond detectors and observe their $\cos\theta_{\mathrm{error}}$ distribution.  Furthermore, we veto series of pixels under the hit position.  Through trial and error, we determine that for the complete rejection of such events, we need to veto not just the pixel immediately underneath the neutron generation vertex, but a wider 13 $\times$ 13 array.  Such a veto ensures that no neutrons produced in the diamond detectors contaminate the final neutron count. This results in significant dead time;  a given EJ-276 pixel is "dead" when an array of 13 $\times$ 13 cm$^2$ is triggered by protons; during this dead time it will miss legitimate neutrons incident from above. We discuss the effects of this dead time on total efficiencies in next section.

\section*{Dead times due to p/e\textsuperscript{-} veto}\label{sec:dead_time}

The harsh space weather environment at the altitude of 2000~km presents multiple challenges.  The first challenge is the dose to the satellite and the electronic instrumentation.  The estimated dose rate in this orbit is $\sim$4 Rad/hr, amounting to approximately 0.7~kRad/week~\cite{mazur2023relativistic}. The radiation-hard sensors, such as EJ-276 scintillators, silicon photomultipliers, and diamond detectors, are optimal for this environment. 

As the protons and electrons impact the CVD diamond outer veto detector, they cause triggers that, once combined in anti-coincidence with the inner EJ-276, create an effective dead time. In other words, every time the diamond veto issues a veto anti-coincidence signal, the associated EJ-276 pixel signal will need to be vetoed.  This could, in principle, cause EJ-276 to miss a legitimate spallation neutron count from the suspect satellite. To estimate the dead times, we consider the fact that EJ-276 has a very fast rise time of $\sim$3~ns and a jitter of FWHM$\approx1$~ns~\cite{pant}.  Neglecting the arrival time of the neutron to the pixel, due to the very short distance, a veto can be applied to the $\pm0.75$~ns within the expected rise of the main EJ-276 pulse. This amounts to a veto window of $\tau\approx1.5$~ns.  If the hit rate is $\nu$ counts per second, and if $\nu \ll 1/\tau$, then the fraction of the time that a given scintillator pixel is vetoed out is $p\approx\nu\tau$.



The dead time calculation is detailed in Methods, yielding a livetime fraction of $p_{\mathrm{live}}\approx0.77$.

\section*{Neutron signal and detection times}\label{sec:signal}

The first step in determining the neutron signal in the 9U inspector is identifying the feasible distance for performing the measurement.  This is primarily a matter of policy:  a flyby is considered too close if it may trigger a political crisis.  There is no hard evidence to claim whether a particular distance is politically acceptable or not.  Nevertheless, there is a significant record of a number of recent examples of US, Chinese, and Russian satellite flybys: see Methods for a detailed listing. Based on these records, we conclude that observation distances of 4-10~km are acceptable for the purposes of this study, in that such past events have taken place without triggering serious political crises. We use $d=4000$~m as a feasible distance for the inspection.

The results of the analysis described in Figure~\ref{fig:time} are used to determine the total neutron count in three orbits.  An integral of neutrons produced by proton spallation yields $Y\approx$9.1$\times$10$^{10}$ neutrons in approximately 6 hrs of observation. We use this source term later to determine the signal in the 9U CubeSat detector.

Having determined an intrinsic efficiency of $\epsilon=0.0054$ and a live time of $p_{\mathrm{live}}=0.77$, we can finally determine the neutron count in a detector of a size 30$\times$30 cm$^2$, from a distance of $d=4000$~m. The mean count is determined using the following formula:
\begin{equation*}
\bar c = \frac{Y}{6}t\frac{0.3^2}{4\pi d^2} \epsilon p_{\mathrm{live}}    ,
\end{equation*}
where  $t$ is the number of hours necessary to achieve this goal. Assuming a Poisson process, one needs $\bar c=5$ to ensure the detection of at least one neutron with probability of $p= \sum_{k=1}^\infty \frac{\bar c^k e^{-\bar c}}{k!} =1-\exp{(-\bar c)}\approx 99\%$. Using $\bar c = 5$ and the equation above, we determine that from the distance of $d=4000$~m the observation time needs to be $t=7.2$~days.

\begin{figure}[!h]
    \centering
    \includegraphics[width=\linewidth]{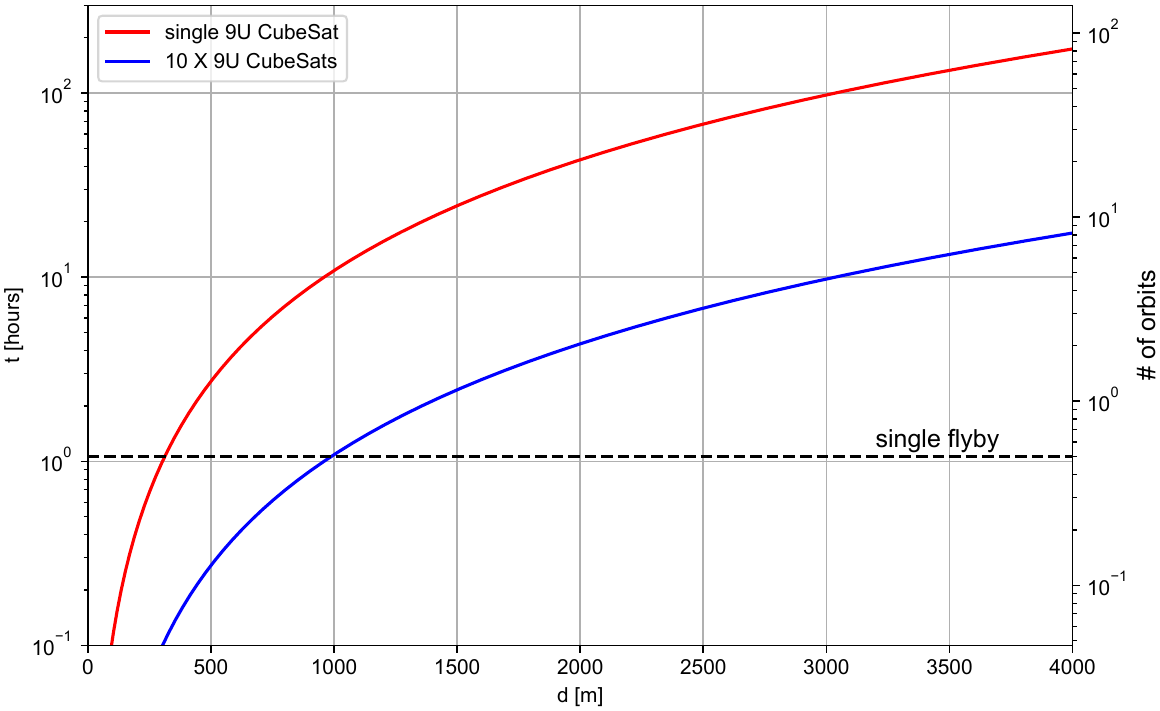}
    \caption{The dependence of the estimated observation time necessary for confirming the presence of a hypothetical thermonuclear device carried by a suspect satellite vs. the measurement distance. The calculation is performed for two scenarios:  a single 9U CubeSat; a constellation of 10 such CubeSats.}
    \label{fig:time_vs_distance}
\end{figure}

To summarize, we conclude that the detection of a thermonuclear weapon placed at an altitude of 2000~km is achievable by a 9U CubeSat inspector from a distance of 4~km with a detection probability of $>99\%$ in approximately one week.  This time can be further reduced to just 15 hours by deploying a constellation of ten 9U CubeSat inspectors. If this constellation is brought to a proximity of 1~km to the suspected satellite, then the signal will increase 16-fold, leading to measurement times of just 1 hour.  This would amount to a single flyby.  The results of this analysis are plotted in Fig.~\ref{fig:time_vs_distance}.  

Simulations were also performed to determine the expected false positive rate in the scenario of a 9U satellite.  The results indicate an upper limit to the false positive rate of $p_\text{fp}<1.1\%$.  See Supplementary Note 3 for further detail. 
This analysis is based on the scenario of unshielded weapons.  Supplementary Note 2 describes a brief discussion and calculations of neutron shielding on the minimum size requirements of the detector. Future studies should further focus on the effects of shielding, as well as the detection of signatures that would be indicative of a shielding attempt. 

Finally, it should be clarified that while proton induced neutron spallation is particularly intense in fissionable materials, it is also present in high-Z elements such as lead and tungsten.  The purpose of this technique is to differentiate a satellite carrying a thermonuclear device from conventional satellites.  These primarily consist of aluminum and hydrogenous materials, which have negligible neutron spallation effects.  This circumstance makes the neutron spallation signal a good indicator of a likely thermonuclear device in the satellite.  

\section*{}

This study has used Geant4 Monte Carlo simulations and empirical models of neutron spallation effects to demonstrate that the confirmation of a satellite carrying a thermonuclear device onboard is scientifically and technologically feasible for an inspector satellite in approximately one week from distances of about 4000 meters, using sensor technologies that are broadly available, such as the EJ-276 PSD-capable plastic scintillator and CVD single crystal diamond detectors. This is by no means a complete or final conclusion.  Future engineering proof of concept studies are needed to test the practicality of this approach. The purpose of this study is to inform policy and provide the theoretical basis for future research in this field. 
As mentioned earlier in the paper, there are many challenges and open questions that need to be addressed for the proposed concept to achieve a high technical readiness level. Some of these are: the high hit rate, harsh radiation environment in the L=1.4 McIlwain L-shell, and its impact on the functioning of sensor components and supporting DAQ electronics; the effects of in-vacuum out-gassing on various sensors, e.g. the EJ-276 scintillator; the impact of heating and cooling on the sensor components and the supporting electronics. This study briefly analyzes the effects of shielding in  Supplementary Note 2; however, more detailed research is necessary for understanding its impact on feasibility. Such studies should also consider the new detection signatures that will be generated, e.g. via (n,$\gamma$) and \textsuperscript{12}C(n,n$^\prime \gamma$)\textsuperscript{12}C reactions.  Future research studies should also focus on producing more precise estimates of the source terms and backgrounds, e.g. by using more detailed proton induced neutron spallation models. Weapon composition will also affect the source term, and future classified research will need to determine its impact on detection times.  Additionally, this study did not address whether it is possible to utilize various valuable features of the space environment, such as the directionality of the proton and electron flux.  In particular, the AP9 model indicates that most protons travel at approximately $90^\circ\pm30^\circ$ to the magnetic field lines.  It may be possible to use this information to significantly simplify the complexity of the proposed detector system, e.g. by forgoing the expensive diamond detectors and rejecting protons based on their direction alone. Finally, the detector array, as described here, can also act as a directional spectrometer for $>$100~MeV protons, possibly allowing the directionality of the proton flux to be used for performing proton radiography of the suspected satellite.

\bmhead{Acknowledgments}  The author would like to thank Grigor Tukharyan for generating the IRENE AP9/AE9 model output used for Fig.~\ref{fig:time}. Special gratitude is due to Prof. Juan Claudio Nino,  Prof. Igor Jovanovic, and Prof. Kyle Hartig for some early discussions about the use of GCR induced signatures. Thanks are due to Dr. Jake Hecla for many valuable discussions. Prof. Vipin Narang, Austin Long, Pranay Vaddi, Greg Ginet, Eric Evans, Brent Parham, Grant Stokes, and Scott Van Broekhoven provided great feedback on the ideas used in this study. The author thanks Dr. Cynthia Nitta from LLNL and Prof. Scott Kemp for introducing him to the problem of OST verification. ChatGPT-5.2 LLM was used for the development of the analysis code, which then underwent thorough testing and verification by the author.  
\bmhead{Funding} {This work was in part funded by the NNSA NA-221 award DE-NA0003920. This publication was made possible in part by a grant from Alfred Carnegie Foundation and Longview Philanthropy USA Inc.. The statements made and views expressed are solely the responsibility of the author.}
\bmhead{Author contributions}  A.D. performed all aspects of the work in this study.
\bmhead{Competing interests} The author declares no competing interests.


\section*{Methods}

This study used the grasshopper application, based on the Geant4 toolkit, to produce a MC simulation model of the interaction of charged particles with the detector~\cite{grasshopper,Agosti-short,Allison2016}.  IRENE AP9 and AE9 models, version 1.58, were used to produce the proton and electron spectra, as described earlier in the text.

\bmhead{Dead layer and Geant4 physics models}
Here, it is tempting to think of the operation of the veto as being based purely on the ionization of the protons in the diamond.  However, there is a significant probability that the 10-100 MeV protons will induce nuclear reactions $^{12}$C(p,nX) in the dead layer and scatter back into the open space. Such reactions will produce no veto signal while generating neutrons, which may trigger events in the EJ-276, thus resulting in a significant background.
This is the reason why the dead layer and the appropriate nuclear reactions are included in the Geant4-based grasshopper simulation~\cite{Agosti-short,Allison2016,grasshopper}. The Monte Carlo (MC) simulations show that these events are, however, insignificant. The Geant4 physics models chosen were {\verb|G4HadronPhysicsQGSP_BIC_HP|} and {\verb|G4HadronElasticPhysicsHP|}.
For the purposes of the grasshopper simulations, the polyvinyl toluene (PVT) scintillator was used as a proxy for EJ-276, due to the near-identity of atomic makeup and the energy deposition responses.

\bmhead{Proton anti-coincidence simulation}
The proton fluxes were studied for the peak of the flux at minute 316 of the orbit, as described in Fig.~\ref{fig:time}(b).  The proton spectrum is fed into the grasshopper MC simulation described earlier.  The simulation is performed for an equivalent fluence of 200 seconds of exposure time.  Every track in the veto and the inner detector is recorded.  In a post-processing step, for every event, the energies of the tracks for each detector are added.  A threshold of $E_{\mathrm{veto}}>0.6$~MeV, which is selected because it is below the minimum deposited energy by a 2~GeV proton, is applied to the energy deposited in the diamond. A $0.2<E_{\mathrm{ej276}}<10$~MeV 
cut is applied in the scintillator, where the energy is in units of MeV proton equivalents (MeVpe). An anti-coincidence logic is then applied between the diamond veto signal and the scintillator signal.  Fig.~\ref{fig:protons_and_theta}(a) histograms the distribution of EJ-276 deposited energy vs. the energy deposited in the diamond veto counters, with the blue patch identifying the acceptance region for neutrons. The results of the analysis of the simulation output show that all proton tracks are rejected via anti-coincidence, ensuring that no proton hits in EJ-276 are mistaken for neutrons. 

\bmhead{Neutron scatter camera kinematics and simulation}
In most neutron scatter camera approaches, the analysis starts with the detected neutron information, and a back-projection cone is calculated, allowing for the determination of the likely location of the source.  In this study, the opposite takes place: we start with the assumption that the neutron came from the suspect located vertically above and then compare the detected scatter angle to the one kinematically calculated based on the deposited energy and time-of-flight information. In this system, the two detection planes are 30 $\times$ 30 arrays of 1~cm pixels, as described earlier, separated by a distance of 10~cm.  This amounts to the size of a 9U CubeSat.   The first plane detects and records the energy $E_\mathrm{p}$ deposited by the recoil proton, as well as the coordinates of the hit with 1~cm precision.  The second plane detects the position and the time difference $\Delta t$.  Assuming a vertically incident neutron and knowing the hit positions, the distance between hits $r^2=\Delta x^2 + \Delta y^2 + \Delta z^2$ and the scattered neutron's  angle $\theta_\mathrm{s}=\arctan [\sqrt{\Delta x^2 + \Delta y^2}/\Delta z]$ are determined.  Finally, the scattered neutron's energy is determined via $E' = m(r/\Delta t)^2/2$.  Knowing $E_\mathrm{p}$ and $E'$ the {\it expected} scatter angle can be determined via $\theta_0 = \arcsin(\sqrt{E_\mathrm{p}/(E'+E_\mathrm{p})})$.  After this, we can determine the error in the scattered angle by comparing the geometric scatter angle and the kinematically reconstructed angle: $\theta_{\mathrm{error}} = \theta_\mathrm{s}-\theta_0$.  For the ideal case $\theta_{\mathrm{error}}=0$.  However, $\theta_\mathrm{error}$ is somewhat smeared due to uncertainties in hit position, energy reconstruction, and time resolution.  Furthermore, the error will have contributions from events in which the recoil occurred on the carbon nuclei. We use $\theta_{\mathrm{error}}$ as the main parameter for differentiating between fissile spallation neutrons coming vertically from above, i.e. from the suspect satellite, and from atmospheric albedo neutrons coming from below.
To study the effectiveness of this approach, a series of Geant4/grasshopper MC simulations were performed, modeling the two planes of neutron detectors in coincidence.  In the post-processing step, resolution effects of the hit coordinates, coincidence times, and energy depositions were applied.  As before, we applied a threshold of $E_\mathrm{p}>0.2$~MeV for recoil proton detection.

\bmhead{Dead time analysis}
In the proposed approach, the diamond veto detector cannot discriminate between electrons and protons. The veto logic must treat electrons and protons identically.  Thus, any charged particle hitting the diamond pixel will require the system to veto the 13$\times$13 array of scintillators underneath, as described earlier, to prevent any neutrons produced via various nuclear reactions, such as $^{12}$C(p,p$'$n)$^{11}$C and $^{12}$C(p,n)$^{12}$N from creating rare neutron coincidences that fall within $\cos\theta_{\mathrm{error}}>0.95$.  In other words, any given EJ-276 at the center of the array will need to be vetoed out any time any of the diamond sensors in the 13$\times$13 array above it are triggered.  At the periphery of the array, this number is, of course, smaller; however, for simplicity, we use the conservative 13$\times$13=169~cm$^{2}$ number for our calculation of the charged particle flux that contributes to the dead time. We then conclude that the fraction of the time a given pixel is vetoed out is $p\approx 169(\phi_\mathrm{e}+\phi_\mathrm{p})\tau=0.23$.  The livetime can be estimated $p_{\mathrm{live}}=1-p\approx0.77$.  This number is used in the calculations of the signal term in the system.

\bmhead{Code Availability}
The analysis of the output was performed using Python scripts and Jupyter notebooks with the numpy, scipy, and matplotlib libraries. The simulations were performed using the grasshopper/Geant4 toolkit~\cite{grasshopper}.  The Geant4 simulation input and macro files, the grasshopper geometry definitions, and the complete Python/Jupyter analysis package used to produce all results and figures in this work are publicly available and can be found in~\url{https://github.com/ustajan/kosmos}. Below is a listing of the most important parts of the modeling toolkit. All the GitHub sub-directories contain a README file with detailed information relevant to the particular directory. 
\bmhead{AP9/AE9 IRENE output}  The output of the IRENE model and a Python script for plotting it are located in~\url{https://github.com/ustajan/kosmos/tree/main/irene/first_space_runs}. These scripts are used to generate Fig.~\ref{fig:time}.
\bmhead{MC model of a single pixel} The Geant4 model of the proton response in a single pixel is located in~\url{https://github.com/ustajan/kosmos/tree/main/protons}.  The directory also contains a subdirectory \verb|postprocessing| with scripts for analyzing the output of the simulation. The scripts and grasshopper input in this directory are used to generate Fig.~\ref{fig:pixel_planes}(a) and~\ref{fig:protons_and_theta}(a).
\bmhead{Neutron directionality analysis} The directory~\url{https://github.com/ustajan/kosmos/tree/main/neutron_backprojection} contains the simulation input files for determining the directional signal of the 9U detector array. It also contains the Jupyter notebooks for analyzing the output of the simulation. The results of this analysis is represented in Fig.~\ref{fig:pixel_planes}(b) and~\ref{fig:protons_and_theta}(b).  The neutron spectrum used in this and other neutron simulations is acquired from Figure 3 of Ref.~\cite{carpenter1977pulsed}, and it can be found in~\url{https://github.com/ustajan/kosmos/blob/main/neutrons/neutron_spallation.tsv}.
\bmhead{Measurement time vs. distance} The final analysis, as described in Section~\nameref{sec:signal} can be found in a Jupyter notebook in~\url{https://github.com/ustajan/kosmos/tree/main/time_vs_distance}.  This analysis is used for generating Fig.~\ref{fig:time_vs_distance}.

\bmhead{Recent history of flybys between the US, Russian, and Chinese satellites}\label{sec:flybys} 
\begin{itemize}
    \item Recently a US surveillance satellite passed a Chinese satellite from the distance of about 12 km. Specifically, the American USA 324 satellite passed by the Chinese satellite TJS-17~\cite{newsweek}.
    \item   In 2022 Russian satellite Luch Olymp flew by US Intelsat 37E at the distance of just 4~km~\cite{kratosspace}.
    \item In 2015 Russian Luch Olymp maneuvered to a distance of less than 10~km to US Intelsat-7 and Intelsat-901~\cite{satnews}.
    \item On February 6, 2025, Luch Olymp 2 approached to Intelsat 1002 to a separation distances less than 5~km\cite{kratosspace2}.
    \item US GSSAP satellites approached to distances of less than 10~km of five Russian GEO-based satellites in the period of 2016-2018, according to Russian space observation data~\cite{breakingdefense}.
    \item PROBA-3 mission is an example of a precise formation-flying technology using two satellites at distances of 150 meters for extended periods of time~\cite{proba}.  
\end{itemize}





\renewcommand\refname{Methods references}%


\end{document}




\section{Supplementary Note: Orbital requirements of an Inspector Satellite}\label{sec:orbit}


A simplified close-proximity inspection scenario was modeled using the linearized Clohessy--Wiltshire (CW) equations about an approximately circular reference orbit. The chief spacecraft is Kosmos2553 with the following orbital parameters at the time of writing of this work: perigee altitude of 2000.2 km; apogee altitude of 2006.6 km; orbital period of  127.1 minutes; inclination of 67.1$^\text{o}$.

From the perigee and apogee radii, the semi-major axis is computed as $ a = \frac{r_p + r_a}{2}$, where $r_p = 2000.2 + r_\text{earth}$ and $r_a = 2006.6 + r_\text{earth}$, and $r_\text{earth}=6371$~km. The period of the mean motion is obtained from Kepler’s third law,
$ n = \sqrt{{\mu}/{a^3}}, T = {2\pi}/{n}$, 
where $\mu=398600~\text{km}^3/\text{s}^2$ is Earth's mean gravitational parameter.

\begin{figure}[pb]
    \centering
    \includegraphics[width=0.95\linewidth]{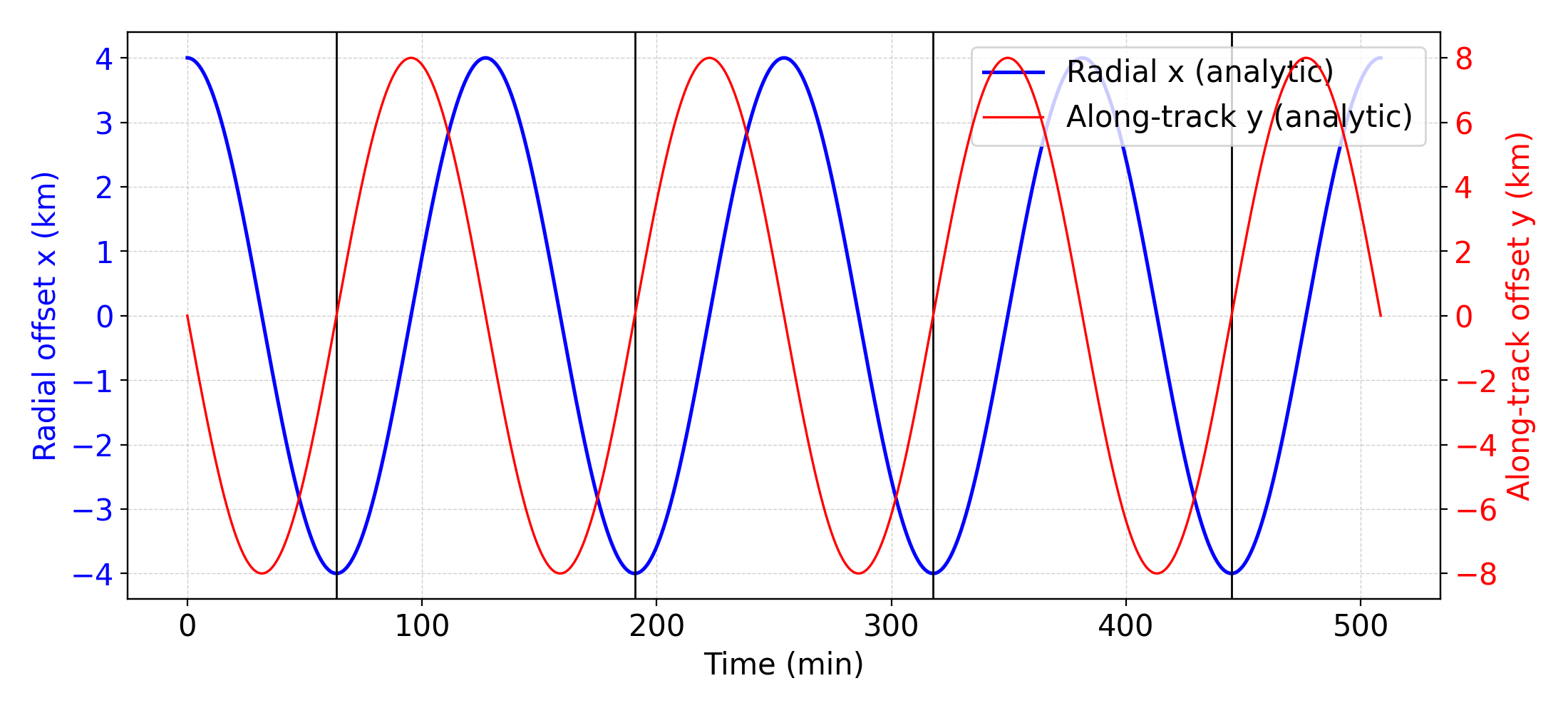}
    \caption{The analytic CW solution plotted over four orbital periods. The radial offset is shown on the left (blue) axis and the along-track offset on the right (red) axis.
Vertical black lines at half-period intervals define the inspection points where the deputy is directly 4~km below the chief. The plot shows that at the time of zero along-track offset the radial offset is $\pm4$~km.}
    \label{fig:relative_motion}
\end{figure}

The deputy spacecraft is an inspector satellite designed to collect data when its radial offset relative to Kosmos2553 equals $-4$ km (i.e., 4 km closer to Earth in the Hill frame). To maintain the same mean motion as the chief, the deputy is assigned a perigee of 1996.2 km, an apogee of 2010.6 km, and the same inclination as the chief. This results in the same exact semi-major axis as the chief and the same orbital period. 
Furthermore, the deputy's phase is selected such that at $t=0$ it is exactly 4~km radially-outward with respect to the chief. 
Given that both the chief and the deputy have very small eccentricities of 
$e = (r_a-r_p)/(r_a+r_p)\approx0$, we neglect eccentricity effects and use the CW circular orbit framework for approximating the relative motion.  The relative motion is described in the Hill frame using the planar CW solutions:
\[
x(t) = A \cos(nt) + B \sin(nt),   ~~~~ y(t) = -2A \sin(nt) + 2B \cos(nt),
\]
where $x$ is the radial offset and $y$ is the along-track offset.
Choosing $A = 4 \text{ km}, B = 0$ produces a bounded relative trajectory in which the radial separation oscillates between $\pm4$~km, while the along-track separation necessarily oscillates between $\pm 8$ km.  The $x(t)$ and $y(t)$ are plotted in Fig.~\ref{fig:relative_motion}.

In addition to achieving relative motion with a $\pm4$~km radial oscillation, the phase of the deputy needs to be chosen to achieve a $-4$~km position relative to the chief when the two cross the equatorial plane, where for $L=1.3$ the Van Allen Radiation belt proton population is at maximum. 

This model suggests that it is possible to design an orbit where the deputy (inspector) can position itself below the chief (Kosmos2553) once per orbit to perform neutron measurements.  
This analysis shows that there is one measurement opportunity per orbit; this can be doubled by having a second inspector with the same parameters but with $A$ chosen to be $A=-4$~km.   The intrinsic coupling of radial and along-track motion in the linear CW dynamics produces a bounded radial oscillation of amplitude $A$ and an along-track oscillation of amplitude $2A$. The resulting periodic geometry defines predictable inspection windows for the deputy spacecraft during each orbit of Kosmos2553. The simple model presented above makes multiple approximations and a more detailed orbital analysis is necessary and should be part of future work. 

\section{Supplementary Note: Weapon composition and neutron source term}

The weapon composition may significantly affect the spallation neutron source term.  Unfortunately, it is impossible to make any exact assumptions about composition, as most thermonuclear weapon design information is classified and publicly inaccessible.  There is  some public information about the US designs that indicates that the mass of the fissionable material on a W87 warhead could be 200 kg (see Ref.~\cite{Harvey01081994}). According to declarations in the framework of the START treaty, the Russian R-36 missile has a throw-weight of 8800 kg and 10 multiple reentry vehicles (MIRV), making the weight of one Russian MIRV 880 kg~\cite{start}. Assuming that the fissile components are the heaviest in the weapon and make up most of its weight, this analysis implies that Russian weapons could be multiple times heavier than W87.  This study assumes only 95 kg, as it is the basis of the Carpenter model used in this study. It is thus reasonable to use the Carpenter model as a basis for neutron spallation as a lower limit for the neutron source term.

Ultimately, the final goal of this study is to build an analysis framework that can be used by specialists with access to classified information to perform more detailed simulations and calculations and thus determine the actual performance of the proposed approach.

\section{Supplementary Note: Shielding effects} 

The study presented in the main body does not include the effects of possible neutron shielding, e.g. by layers of polyethylene, which the designers of the suspected satellite could use to reduce the neutron source term.  Understanding what shielding types are possible for a weapon placed in the outer space environment is not a simple task, which is why it has been delegated to future research efforts. While in ground-based systems neutron shielding is trivial, in outer space it is a much more complex problem. The neutron shielding will  act as thermal insulation, which will make it more difficult to remove the $\sim$10~W of alpha heating and any additional heating due to the increase in fission triggered by protons.  Given that the weapon is expected to serve for years, any accumulation of heat due to the additional insulation may result in an increase in the weapon component temperatures, possibly affecting their reliability in the case of nuclear use. 

Ignoring these circumstances, it is possible to estimate the impact of the shielding on the required size of the inspector satellite for various thicknesses of polyethylene shielding.  Polyethylene, being hydrogen-rich, moderates the fast $\sim$2 MeV neutrons produced by proton spallation out of the energy acceptance of the scintillator. A series of grasshopper/Geant4 simulations were performed, where an isotropic neutron source was surrounded by varying thicknesses of polyethylene, and the counts of a nominal detector with energies of $E_\text{dep}>0.2$~MeV were used to estimate the transmission $T=c(x)/c(0)$, where $c(x)$ are the counts for shielding of thickness $x$. The neutron energies were sampled from the spectrum presented in Ref.~\cite{carpenter1977pulsed}. 

To maintain the goal of detection in one week from a distance of 4~km in scenarios with shielding, the size of the CubeSat needs to be increased by a factor proportional to $1/T$ from 9U to XU, where X can be determined from $X = 9/T$.
Table~\ref{tab:shielding} lists the various possible values of shielding thickness and the corresponding required sizes of the CubeSat. For example, for 20~cm of polyethylene shielding, the inspector needs to have the form factor of at least 265U, i.e. make up a panel of approximately 1.6$\times$1.7 m$^2$.
\begin{table}[h]
\centering
\begin{tabular}{cccc}
\hline
$x$ (cm) & $T$ & minimum $\lceil X \rceil$ & detector size (m$\times$m) \\
\hline
0  & 1     & 9   & 0.3 $\times$ 0.3 \\
5  & 0.3   & 30  & 0.6 $\times$ 0.5 \\
10 & 0.12  & 75  & 0.9 $\times$ 0.9 \\
20 & 0.034 & 265 & 1.6 $\times$ 1.7 \\
\hline
\end{tabular}
\caption{Values of CubeSat $X$ and its required size for various values of shielding thickness to achieve a detection from the distance of 4~km in one week. The detector size is rounded to the nearest upper decimeter.}\label{tab:shielding}
\end{table}

\section{Supplementary Information:  False Positive Rate}

In this study, the expected spallation neutron count for a 9U satellite at a distance of 4~km, assuming approximately one week of observation time, has been five.  With a threshold for detection of $c=1$ it can be determined that the probability of detection is $p=1-\exp(-5)=99.3\%$.  
The false positive rate is driven by multiple possible sources.  The greatest of these is the mixing of atmospheric neutrons arriving from below the inspector and spallation neutrons arriving from above.  The main body of the manuscript describes in detail the use of the neutron scatter camera technique by calculating $\theta_\text{error}$  and applying $\cos\theta_\text{error}>0.95$ to achieve very efficient rejection of atmospheric neutrons.   To determine the likelihood of an atmospheric neutron being tagged as a spallation, we ran a longer version of the simulation described in Fig 3(b) in the main body.  The simulation sampled 1.68~bln neutrons and counted those that fell into the $\cos\theta_\text{error}>0.95$,  determining that the number of such counts is zero. We thus conclude that $c_\text{error}<1$ per 1.68~bln incident neutrons and use this as an upper limit for the false positive rate.
Analysis of the atmospheric and albedo neutron background at the altitude of 2000~km, using the model provided in Ref.~\cite{cumani2019background}, yields a rate of 0.033 cm$^{-2}$s$^{-1}$ at the altitude of 2000~km and latitude of 0$^\circ$. 
Using the $t=7.2$~day observation time with a 9U satellite, we can then determine the expectation of false spallation neutrons. That is $\bar c_\text{false} < (1/1.68\times 10^9)\cdot 0.033\cdot (30^2)\cdot 7.2\cdot 24 \cdot 3600 = 0.011 $.  We can then determine the probability of a false positive during this measurement period.  Using the $c_\text{threshold}=1$, is $p_\text{fp}<1-\exp(-0.011)=0.011=1.1\%$.  We thus conclude that, based on the MC simulations performed for the primary scenario described in this study, the false positive probability is $<1.1\%$.

\section*{Supplementary References}
\bibliography{sn-bibliography}